\documentclass[12pt]{article}

\usepackage{graphicx}
\begin{document}

\begin{center}
{\bf Einstein-AdS gravity coupled to nonlinear electrodynamics, magnetic black holes, thermodynamics in an extended phase space and Joule--Thomson expansion} \\
\vspace{5mm} S. I. Kruglov
\footnote{E-mail: serguei.krouglov@utoronto.ca}
\underline{}
\vspace{3mm}

\textit{Department of Physics, University of Toronto, \\60 St. Georges St.,
Toronto, ON M5S 1A7, Canada\\
Canadian Quantum Research Center, \\
204-3002 32 Ave Vernon, BC V1T 2L7, Canada} \\
\vspace{5mm}
\end{center}
\begin{abstract}

We study Einstein's gravity with negative cosmological constant coupled to nonlinear electrodynamics proposed earlier.
The metric and mass functions and corrections to the Reissner--Nordstr\"{o}m solution are obtained. Black hole solutions can have one or two horizons. Thermodynamics and phase transitions of magnetically charged black holes in Anti-de Sitter spacetime are investigated.
The first law of  black hole thermodynamics is formulated and the generalized Smarr relation is proofed. By calculating the Gibbs free energy and heat capacity we study the black hole stability. Zero-order (reentrant), first-order, and second-order phase transitions are analysed.
The Joule--Thomson expansion is considered showing the cooling and heating phase transitions. It was shown that the principles of causality and unitarity are satisfied in the model under consideration.
\end{abstract}


\section{Introduction}

Black holes behave as the thermodynamic systems \cite{Bardeen,Jacobson,Padmanabhan} and they have the entropy connected with the surface area and the surface gravity defines the temperature \cite{Bekenstein,Hawking}. Black holes phase transitions occur in Anti-de Sitter (AdS) spacetime, where the cosmological constant is negative \cite{Page}. It was discovered that gravity in AdS spacetime is linked with the conformal field theory (the holographic principle) \cite{Maldacena} that has an application in condensed matter physics.
In black hole thermodynamics (in an extended phase space) the negative cosmological constant being a thermodynamic pressure which is conjugated to a black hole volume \cite{Dolan,Kubiznak,Mann,Teo}. In Einstein-AdS gravity coupled to nonlinear electrodynamics (NED) {with coupling $\beta$) the constant $\beta$ is conjugated to the vacuum polarization.
The first NED was proposed by M. Born and L. Infeld \cite{Born} to remove a singularity of a point charge and to have the finite the electromagnetic field energy. At the weak-field limit Born--Infeld electrodynamics becomes Maxwell’s theory.
Another NED model was formulated by W. Heisenberg and H. Euler \cite{Heisenberg}, where nonlinearity is due to creation of the electron-positron pairs within quantum electrodynamics. The interest in NED, as a source of gravity, is because of the possibility
to have regular black holes and soliton-like configurations without singularities.
Recent reviews of NED models were given in \cite{Sorokin,Bronnikov0}.
Black hole thermodynamics in Einstein-AdS gravity coupled to Born--Infeld electrodynamics was considered in \cite{Fernando,Dey,Cai,Fernando1,Myung,Banerjee,Miskovic} (see also \cite{Pourhassan1, Pourhassan2}). Born--Infeld-AdS thermodynamics of black holes in extended phase space was studied in \cite{Mann1,Zou,Hendi,Hendi1,Zeng}.
The Joule--Thomson expansion of black holes was investigated in \cite{Aydiner,Yaraie,Rizwan,Chabab,Mirza,Kruglov,Kruglov1,Kruglov2,Kruglov3,Kruglov4}.
In this paper we study a modified Einstein-AdS theory with a NED model, as a matter field, to smooth out singularities of the linear Maxwell theory. We use NED theory with Lagrangian of the form
${\cal L}(\mathcal{F}) =-{\cal F}/\left(4\pi\left(1+(2\beta{\cal F})^{3/4}\right)\right)$, where
${\cal F}=F^{\mu\nu}F_{\mu\nu}/4$ with $F_{\mu\nu}$ being the electromagnetic field tensor.  The interest in this model is due to its simplicity, the metric and mass functions are expressed in the form of elementary functions but in Born--Infeld NED they are special functions.
We consider magnetically charged black holes because electrically charged black holes with NED possessing weak-field Maxwell limit have a singularity \cite{Bronnikov}.
It is worth mentioning that Lagrangians of NED models in the weak-field limit are different. This leads to different
indexes of diffraction and birefringent effects. The similarities in the behavior of critical isotherms, the magnetic potential, vacuum
polarization, the Gibbs free energy, and heat capacity take place for Einstein-AdS gravity coupled to NED models.
Here, an attention is paid on gravity in the AdS (not in de Sitter) spacetime because this case allows us to introduce a pressure which is necessary to consider an extended phase space and thermodynamics. In addition, only in this case the holographic principle occurs.

In section 2 we obtain the metric function and its asymptotic with corrections to the Reissner--Nordstr\"{o}m solution. The first law of black hole thermodynamics in the extended phase space is studied in section 3. We calculate the thermodynamic magnetic potential and the thermodynamic conjugate to the NED coupling (the vacuum polarization). We show that the generalized Smarr relation holds. In section 4, the critical temperature and critical pressure are obtained. By analysing the Gibbs free energy and heat capacity we show that phase transitions take place. It is demonstrated that black hole thermodynamics is similar to Van der Waals thermodynamics. We analyse first-order, second-order, and reentrant phase transitions. The Joule--Thomson adiabatic expansion is studied in section 5. The Joule--Thomson coefficient and the inversion temperature are calculated. Section 6 is a summary. In Appendix A we calculate the Kretschmann scalar. We study causality and unitarity of our NED model in  Appendix B. We show that the principles of causality and unitarity take place for any magnetic induction fields.

We use the units: $c=\hbar=1$, $k_B=1$.

\section{Einstein-AdS black hole solution}

The Einstein-AdS action with the matter is given by
\begin{equation}
I=\int d^{4}x\sqrt{-g}\left(\frac{R-2\Lambda}{16\pi G_N}+\mathcal{L}(\mathcal{F}) \right),
\label{1}
\end{equation}
where $\Lambda=-3/l^2$ is negative cosmological constant with $l$ being the AdS radius. Here, we use the matter Lagrangian in the form of  NED \footnote{We insert the factor $4\pi$ in the denominator of Eq. (2) to use the Gaussian units compared to Heaviside--Lorentz units explored in \cite{Kruglov5}.} \cite{Kruglov5}
\begin{equation}
{\cal L}(\mathcal{F}) =-\frac{{\cal F}}{4\pi\left(1+(2\beta{\cal F})^{3/4}\right)},
\label{2}
\end{equation}
with ${\cal F}=F^{\mu\nu}F_{\mu\nu}/4=(B^2-E^2)/2$, where $E$ and $B$ are the electric and magnetic induction fields, respectively. As $\beta\rightarrow 0$ Lagrangian (2) becomes the Maxwell Lagrangian ${\cal L}_M=-\mathcal{F}/(4\pi)$.
From action (1) one obtains the field equations
\begin{equation}
R_{\mu\nu}-\frac{1}{2}g_{\mu \nu}R+\Lambda g_{\mu \nu} =8\pi G_N T_{\mu \nu},
\label{3}
 \end{equation}
\begin{equation}
\partial _{\mu }\left( \sqrt{-g}\mathcal{L}_{\mathcal{F}}F^{\mu \nu}\right)=0,
\label{4}
\end{equation}
where $\mathcal{L}_{\mathcal{F}}=\partial \mathcal{L}( \mathcal{F})/\partial \mathcal{F}$. The energy-momentum tensor reads
\begin{equation}
 T_{\mu\nu }=F_{\mu\rho }F_{\nu }^{~\rho }\mathcal{L}_{\mathcal{F}}+g_{\mu \nu }\mathcal{L}\left( \mathcal{F}\right).
\label{5}
\end{equation}
The line element squared with spherical symmetry is
\begin{equation}
ds^{2}=-f(r)dt^{2}+\frac{1}{f(r)}dr^{2}+r^{2}\left( d\theta
^{2}+\sin ^{2}(\theta) d\phi ^{2}\right).
\label{6}
\end{equation}
We treat the black hole as a magnetic monopole with the magnetic induction field $B=q/r^2$, where $q$ is the magnetic charge.
The metric function is given by \cite{Bronnikov}
\begin{equation}
f(r)=1-\frac{2m(r)G_N}{r},
\label{7}
\end{equation}
with the mass function
\begin{equation}
m(r)=m_0+4\pi\int_{0}^{r}\rho (r)r^{2}dr.
\label{8}
\end{equation}
Here, $m_0$ is an integration constant (the Schwarzschild mass), and $\rho$ is the energy density.
Making use of Eq. (5) the magnetic energy density plus the energy density due to AdS spacetime is given by
\begin{equation}
\rho=\frac{q^2}{8\pi r\left(r^3+(\beta q^2)^{3/4}\right)}-\frac{3}{8\pi G_Nl^2}.
\label{9}
\end{equation}
From Eqs. (8) and (9) one obtains the mass function
\[
m(r)=m_0+\frac{q^{3/2}}{12\sqrt[4]{\beta }}\biggl[
\ln\frac{r^2-\sqrt[4]{\beta q^2}r+\sqrt{\beta} q}{(r+\sqrt[4]{\beta q^2})^2}
\]
\begin{equation}
-2\sqrt{3}\arctan\left(\frac{1-2r/\sqrt[4]{\beta q^2}}{\sqrt{3}}\right)+\frac{\pi}{\sqrt{3}}\biggr]-\frac{r^3}{2G_Nl^2}.
\label{10}
\end{equation}
The magnetic energy of the black hole becomes
\begin{equation}
m_M=\frac{q^2}{2}\int_0^\infty \frac{r}{r^3+(\beta q^2)^{3/4}}dr=\frac{\pi q^{3/2}}{3\sqrt{3}\sqrt[4]{\beta}}.
\label{11}
\end{equation}
The magnetic energy, which can be considered as a magnetic mass, is finite. Thus, the coupling $\beta$ smoothes singularities. Making use of Eqs. (7) and (10) we obtain the metric function
\[
f(r)=1-\frac{2m_0 G_N}{r}-\frac{q^{3/2}G_N}{6\sqrt[4]{\beta}r}\biggl[
\ln\frac{r^2-\sqrt[4]{\beta q^2}r+\sqrt{\beta} q}{(r+\sqrt[4]{\beta q^2})^2}
\]
\begin{equation}
-2\sqrt{3}\arctan\left(\frac{1-2r/\sqrt[4]{\beta q^2}}{\sqrt{3}}\right)+\frac{\pi}{\sqrt{3}}\biggr]+\frac{r^2}{l^2}.
\label{12}
\end{equation}
As $r\rightarrow 0$, when the Schwarzschild mass is zero ($m_0=0$), one finds
\begin{equation}
f(r)=1-\frac{G_N\sqrt{q}r}{2\beta^{3/4}}+\frac{r^2}{l^2}+\frac{G_Nr^4}{5\beta^{3/2}q}+{\cal O}(r^6).
\label{13}
\end{equation}
As a result, we have $f(0)=1$. The finiteness of the metric function is necessary condition in order to have the spacetime regular. But the spacetime singularity is present in the model because the Kretschmann scalar becomes infinite at $r=0$ (see Appendix A).
Making use of Eq. (12) (when $\Lambda=0$) as $r\rightarrow \infty$, we obtain
\begin{equation}
f(r)=1-\frac{2MG_N}{r}+\frac{q^2G_N}{r^2}-\frac{q^{7/2}\beta^{3/4} G_N}{4r^5}+\mathcal{O}(r^{-6}).
\label{14}
\end{equation}
We define $M=m_0+m_M$ being the ADM mass. According to Eq. (14) black holes have corrections to the Reissner--Nordstr\"{o}m solution. When $\beta=0$ the metric (14) becomes the Reissner--Nordstr\"{o}m metric. The plot of metric function (12) is depicted in Fig. 1 (at $m_0=0$, $G_N=1$, $l=10$).
\begin{figure}[h]
\includegraphics {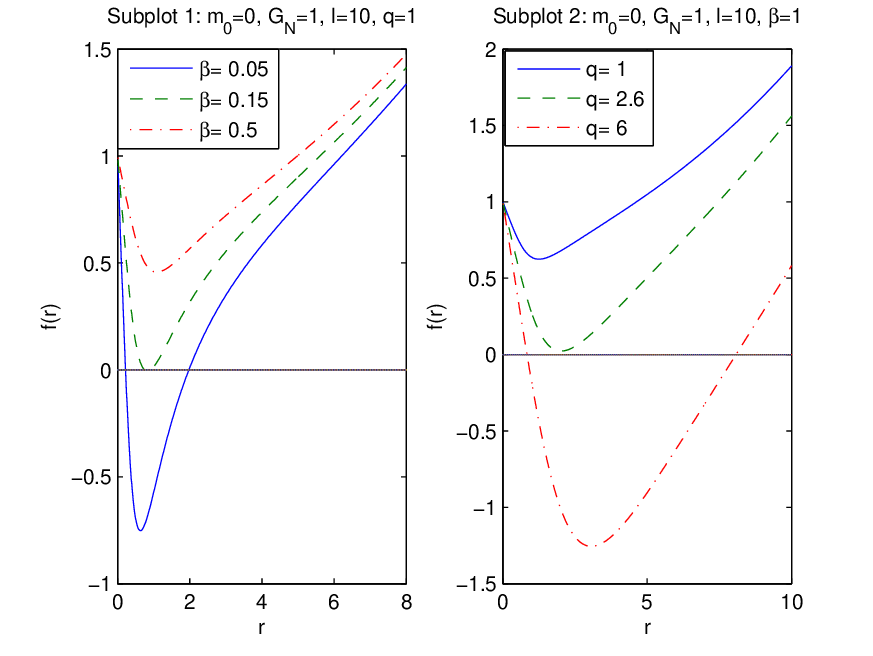}
\caption{\label{fig.1} The function $f(r)$ at $m_0=0$, $G_N=1$, $l=10$. Figure 1 shows that black holes could have one or two horizons. In accordance with subplot 1, if coupling $\beta$ is increasing the event horizon radius decreases. According to subplot 2  when magnetic charge $q$ increases the event horizon radius also increases.}
\end{figure}
According to Fig. 1 black holes may have one or two horizons. When coupling $\beta$ increases, at the constant $q$, the event horizon radius is decreasing. If magnetic charge $q$ increases, at the constant $\beta$, the event horizon radius also increases.

\section{First law of black hole thermodynamics}

We will consider the first law of black hole thermodynamics in extended phase space, where the pressure is $P=-\Lambda/(8\pi)$  \cite{Kastor,Dolan1,Cvetic,Kubiznak1,Kubiznak2} and coupling $\beta$ is the thermodynamic value. In this approach mass $M$ is a chemical enthalpy ($M=U+PV$ with $U$ being the internal energy). By using the Euler's dimensional analysis with $G_N=1$  \cite{Smarr}, \cite{Kastor},
we obtain dimensions as follows: $[M]=L$, $[S]=L^2$, $[P]=L^{-2}$, $[J]=L^2$, $[q]=L$, $[\beta]=L^2$.  Then one finds
\begin{equation}
M=2S\frac{\partial M}{\partial S}-2P\frac{\partial M}{\partial P}+2J\frac{\partial M}{\partial J}+q\frac{\partial M}{\partial q}+2\beta\frac{\partial M}{\partial \beta},
\label{15}
\end{equation}
where $J$ is black hole angular momentum.
The thermodynamic conjugate to coupling $\beta$ is the vacuum polarization \cite{Teo} ${\cal B}=\partial M/\partial \beta $. The black hole entropy $S$, volume $V$ and pressure $P$ are defined as
\begin{equation}
S=\pi r_+^2,~~~V=\frac{4}{3}\pi r_+^3,~~~P=-\frac{\Lambda}{8\pi}=\frac{3}{8\pi l^2}.
\label{16}
\end{equation}
Making use of Eq. (12) for non-rotating black holes ($J=0$) we obtain
\[
M(r_+)=\frac{r_+}{2G_N}+\frac{r_+^3}{2G_Nl^2}+\frac{\pi q^{3/2}}{4\sqrt{3}\sqrt[4]{\beta}}-\frac{q^{3/2}g(r_+)}{12\sqrt[4]{\beta}},
\]
\begin{equation}
g(r_+)=\ln\frac{r_+^2-\sqrt[4]{\beta q^2}r_++\sqrt{\beta} q}{(r_++\sqrt[4]{\beta q^2})^2}
-2\sqrt{3}\arctan\left(\frac{1-2r_+/\sqrt[4]{\beta q^2}}{\sqrt{3}}\right),
\label{17}
\end{equation}
where $r_+$ is the event horizon radius, $f(r_+)=0$.
The total black hole mass $M(r_+)$ versus $r_+$ is plotted in Fig. 2.
\begin{figure}[h]
\includegraphics {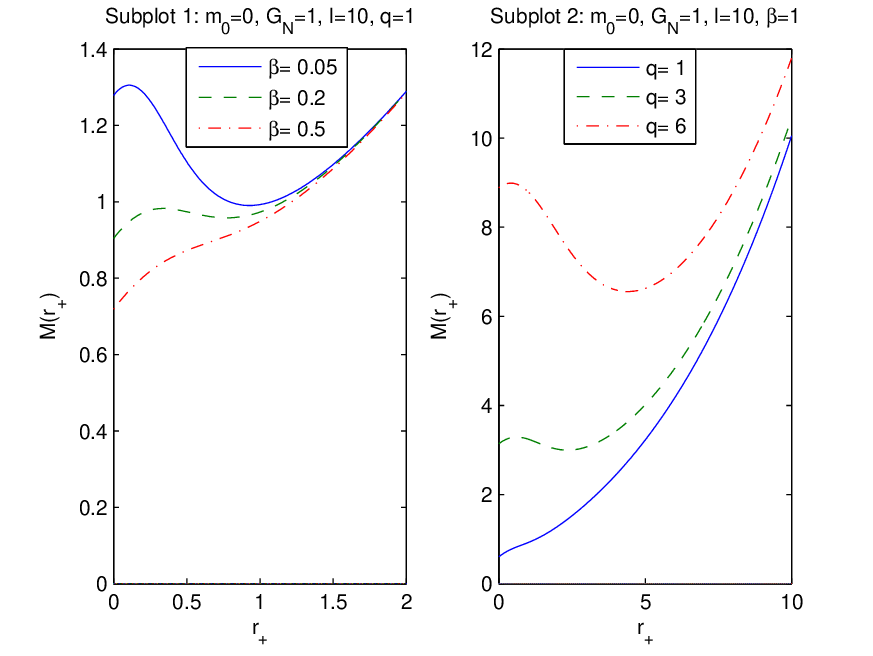}
\caption{\label{fig.2} The function $M(r_+)$ at $m_0=0$, $G_N=1$. According to Fig. 2, left panel, the black hole mass $M(r_+)$ decreases, at fixed $r_+$ and $q$, when the coupling $\beta$ increases. In accordance with right panel, when magnetic charge $q$ increases, at fixed $r_+$ and $\beta$, the event horizon radius also increases.}
\end{figure}
With the help of Eq. (17) we find
\[
dM(r_+)=\left[\frac{1}{2}+\frac{3r_+^2}{2l^2}-\frac{q^2r_+}{2(r_+^3+(\beta q^2)^{3/4})}\right]dr_+
-\frac{r_+^3}{l^3}dl
\]
\[
+\left[-\frac{\sqrt{q}g(r_+)}{8\beta^{1/4}}+\frac{\sqrt{3q}\pi}{8\beta^{1/4}}+\frac{qr_+^2}{4[r_+^3+(\beta q^2)^{3/4}]}\right]dq
\]
\begin{equation}
+\left[\frac{q^{3/2}g(r_+)}{48\beta^{5/4}}-\frac{q^{3/2}\pi}{16\sqrt{3}\beta^{5/4}}+\frac{q^2r_+^2}{8\beta[r_+^3+(\beta q^2)^{3/4}]}\right]d\beta.
\label{18}
\end{equation}
The Hawking temperature is given by
\begin{equation}
T=\frac{f'(r)|_{r=r_+}}{4\pi},
\label{19}
\end{equation}
where $f'(r)=\partial f(r)/\partial r$. By virtue of Eqs. (12) and (19), one obtains the Hawking temperature
\begin{equation}
T=\frac{1}{4\pi}\biggl[\frac{1}{r_+}+\frac{3r_+}{l^2}-\frac{q^2}{r_+^3+(\beta q^2)^{3/4}}\biggr].
\label{20}
\end{equation}
Equation (20) is converted into the Hawking temperature of Maxwell-AdS black hole as $\beta\rightarrow 0$.
Making use Eqs. (15), (18) and (20) we find the first law of black hole thermodynamics
\begin{equation}
dM = TdS + VdP + \Phi dq + {\cal B}d\beta.
\label{21}
\end{equation}
Comparing Eq. (18) with (21) we obtain the magnetic potential $\Phi$ and the vacuum polarization ${\cal B}$
\[
\Phi =-\frac{\sqrt{q}g(r_+)}{8\beta^{1/4}}+\frac{\sqrt{3q}\pi}{8\beta^{1/4}}+\frac{qr_+^2}{4[r_+^3+(\beta q^2)^{3/4}]},
\]
\begin{equation}
{\cal B}=\frac{q^{3/2}g(r_+)}{48\beta^{5/4}}-\frac{q^{3/2}\pi}{16\sqrt{3}\beta^{5/4}}+\frac{q^2r_+^2}{8\beta[r_+^3+(\beta q^2)^{3/4}]}.
\label{22}
\end{equation}
The plots of $\Phi$ and ${\cal B}$ vs $r_+$ are depicted in Fig. 2.
\begin{figure}[h]
\includegraphics {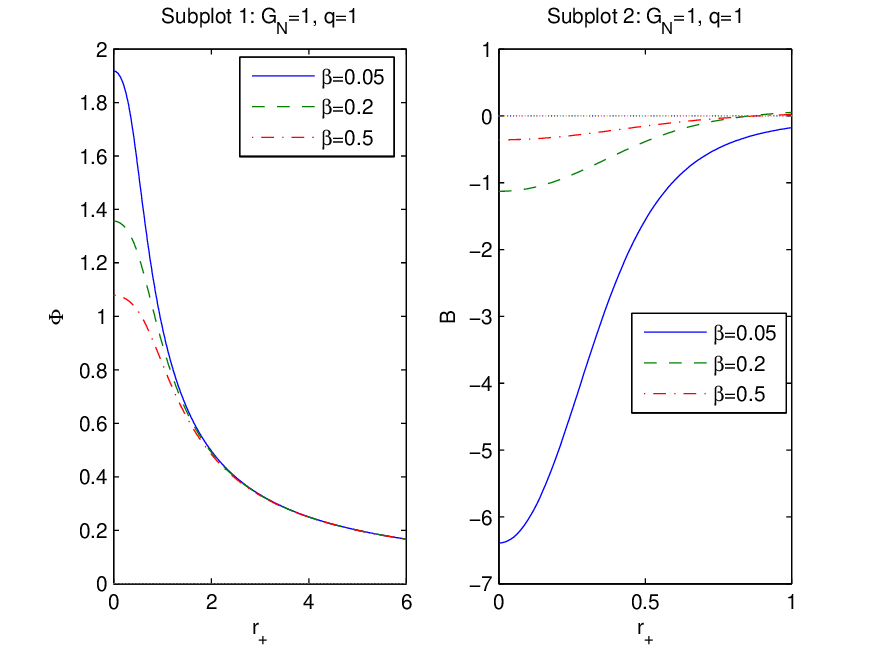}
\caption{\label{fig.3} The functions $\Phi$ and ${\cal B}$ vs $r_+$ at $q=1$. The solid curve in subplot 1 is for $\beta=0.05$, the dashed curve is for $\beta=0.2$, and the dashed-doted curve is for $\beta=0.5$. It follows that the magnetic potential $\Phi$ is finite at $r_+=0$ and becomes zero at $r_+\rightarrow \infty$. If coupling $\beta$ is increasing the magnetic potential decreases. The function ${\cal B}$ in subplot 2 vanishes as $r_+\rightarrow \infty$ and is finite at $r_+=0$. }
\end{figure}
According to Fig. 3 (subplot 1) when parameter $\beta$ increases the magnetic potential $\Phi$ is decreasing. As $r_+\rightarrow \infty$ the magnetic potential vanishes ($\Phi(\infty)=0$), but at $r_+ = 0$ $\Phi$ is finite.
Figure 3 (subplot 2) shows that at $r_+ = 0$ the vacuum polarization is finite and when $r_+\rightarrow \infty$, ${\cal B}$ is zero (${\cal B}(\infty)=0$).

Making use of Eqs. (15), (16) and (22) one can verify that the generalized Smarr relation
\begin{equation}
M=2ST-2PV+q\Phi+2\beta{\cal B}
\label{23}
\end{equation}
holds.
\section{Thermodynamics of black hole }

With the help of Eq. (20) one finds the black hole equation of state
\begin{equation}
P=\frac{T}{2r_+}-\frac{1}{8\pi r_+^2}+\frac{q^2}{8\pi r_+[r_+^3+(\beta q^2)^{3/4}]}.
\label{24}
\end{equation}
Equation (24), as $\beta\rightarrow 0$, is converted into charged  Maxwell-AdS black hole equation of state \cite{Kubiznak1}.
Equation (24) is similar to the Van der Waals equation of state if the specific volume reads $v=2l_Pr_+$ ($l_P=\sqrt{G_N}=1$) \cite{Kubiznak1}. Then Eq. (24) becomes
\begin{equation}
P=\frac{T}{v}-\frac{1}{2\pi v^2}+\frac{2q^2}{\pi v[v^3+8(\beta q^2)^{3/4}]}.
\label{25}
\end{equation}
The inflection points in the $P-v$ diagrams (critical points) may be obtained by equations
\[
\frac{\partial P}{\partial v}=-\frac{T}{v^2}+\frac{1}{\pi v^3}-\frac{8q^2[v^3+2(q^2\beta)^{3/4}]}{\pi v^2[v^3+8(\beta q^2)^{3/4}]^2}=0,
\]
\begin{equation}
\frac{\partial^2 P}{\partial v^2}=\frac{2T}{v^3}-\frac{3}{\pi v^4}+\frac{8q^2[5v^6+8(\beta q^2)^{3/4}v^3+32(\beta q^2)^{3/2}]}{\pi v^3[v^3+8(\beta q^2)^{3/4}]^3}=0.
\label{26}
\end{equation}
By virtue of Eq. (26) one finds the critical points equation
\begin{equation}
[v_c^3+8(\beta q^2)^{3/4}]^3-24q^2v_c^4[v_c^3-4(\beta q^2)^{3/4}]=0.
\label{27}
\end{equation}
Making use of Eq. (26) we obtain the critical temperature and pressure
\begin{equation}
T_c=\frac{1}{\pi v_c}-\frac{8q^2[v^3+2(\beta q^2)^{3/4}]}{\pi [v^3+8(\beta q^2)^{3/4}]^2},
\label{28}
\end{equation}
\begin{equation}
P_c=\frac{1}{2\pi v_c^2}-\frac{6q^2v_c^2}{\pi (v_c^3+8(\beta q^2)^{3/4})^2}.
\label{29}
\end{equation}
The solutions (approximate) $v_c$ to Eq. (27), critical temperatures $T_{c}$ and pressures $P_{c}$ are presented in Table 1.
\begin{table}[ht]
\caption{Critical values of the specific volume, temperatures  and pressures at $q=1$}
\centering
\begin{tabular}{c c c c c c c c c}\\[1ex]
\hline
$\beta$ & 0.1 & 0.2  & 0.4 & 0.5 & 0.7 & 0.8 & 0.9 & 1\\[0.5ex]
\hline
$v_{c}$ &4.790 & 4.708 & 4.552 & 4.472 & 4.297 & 4.196 & 4.080 & 3.936\\[0.5ex]
\hline
$T_{c}$ &0.0438 & 0.0442 & 0.0448 & 0.0452  & 0.0459 & 0.0463 & 0.0467 & 0.0472\\[0.5ex]
\hline
$P_{c}$ &0.0034  & 0.0035 & 0.0036 & 0.0037 & 0.0038 & 0.0039 & 0.0040  & 0.0041\\[0.5ex]
\hline
\end{tabular}
\end{table}
The $P-v$ diagrams are given in Fig.4.
\begin{figure}[h]
\includegraphics {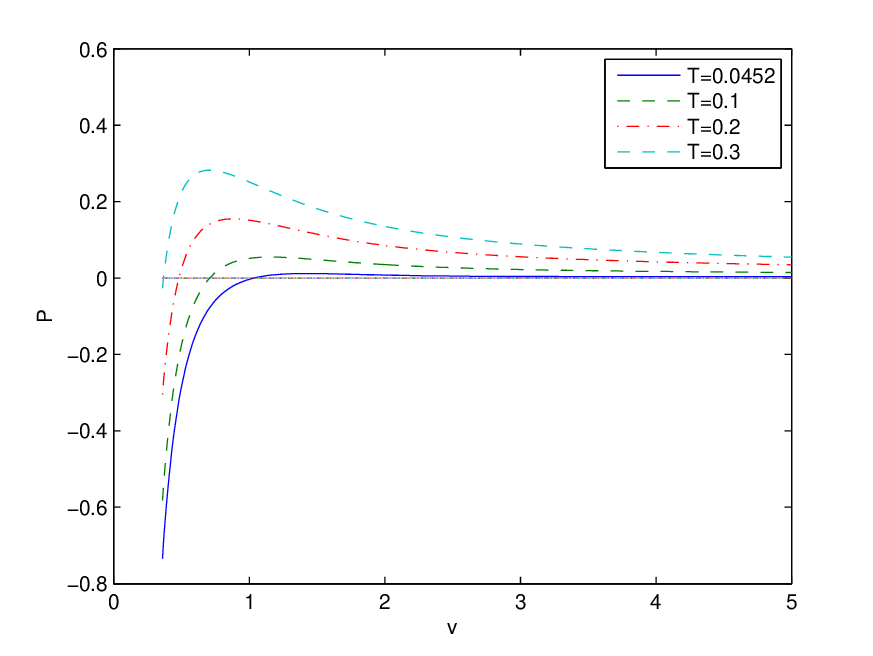}
\caption{\label{fig.4} The function $P(v)$ at $q=1$, $\beta=0.5$. The critical isotherm corresponds to $T_{c}\approx 0.0452$ possessing the inflection point.}
\end{figure}
At $q=1$, $\beta=0.5$ the critical specific volume is $v_{c}\approx 4.472$ and the critical temperature is $T_{c}=0.0452$.
Figure 4 shows that at some point the pressure is zero corresponding to the black hole remnant. Then if the specific volume is increasing the pressure increases and the pressure has a maximum. Then the pressure decreases that is similar to ideal gas. At the critical values we have similarities with Van der Waals liquid behavior having the inflection point.
Making use of Eqs. (27), (28) and (29) and for small $\beta$ one finds
\begin{equation}
v_c^2=24q^2+{\cal O}(\beta),~~~T_c=\frac{1}{3\sqrt{6}\pi q}+{\cal O}(\beta),~~~
P_c=\frac{1}{96\pi q^2}+{\cal O}(\beta).
\label{30}
\end{equation}
At $\beta=0$ in Eq. (30), we obtain the critical points of charged AdS black hole \cite{Mann1}. Then the critical ratio becomes
\begin{equation}
\rho_c=\frac{P_cv_c}{T_c}=\frac{3}{8}+{\cal O}(\beta),
\label{31}
\end{equation}
with the value $\rho_c=3/8$ corresponding to the Van der Waals fluid.

The Gibbs free energy for fixed charge $q$, coupling $\beta$ and pressure $P$ is given by
\begin{equation}
G=M-TS,
\label{32}
\end{equation}
where $M$ is considered as a chemical enthalpy. Making use of Eqs. (16), (17), (20) and (32) we obtain
\begin{equation}
G=\frac{r_+}{4}-\frac{2\pi r_+^3P}{3}+\frac{\pi q^{3/2}}{4\sqrt{3}\beta^{1/4}}+\frac{q^2r_+^2}{4[r_+^3+(\beta q^2)^{3/4}]}
-\frac{q^{3/2}g(r_+)}{12\beta^{1/4}}.
\label{33}
\end{equation}
The plot of the Gibbs free energy $G$ versus $T$ for $\beta=0.5$ and $v_c\approx 4.472$, $T_c\approx 0.0452$ is depicted in Fig. 5. We took into consideration that $r_+$ is the function of $P$ and $T$ (see  Eq. (24)).
\begin{figure}[h]
\includegraphics {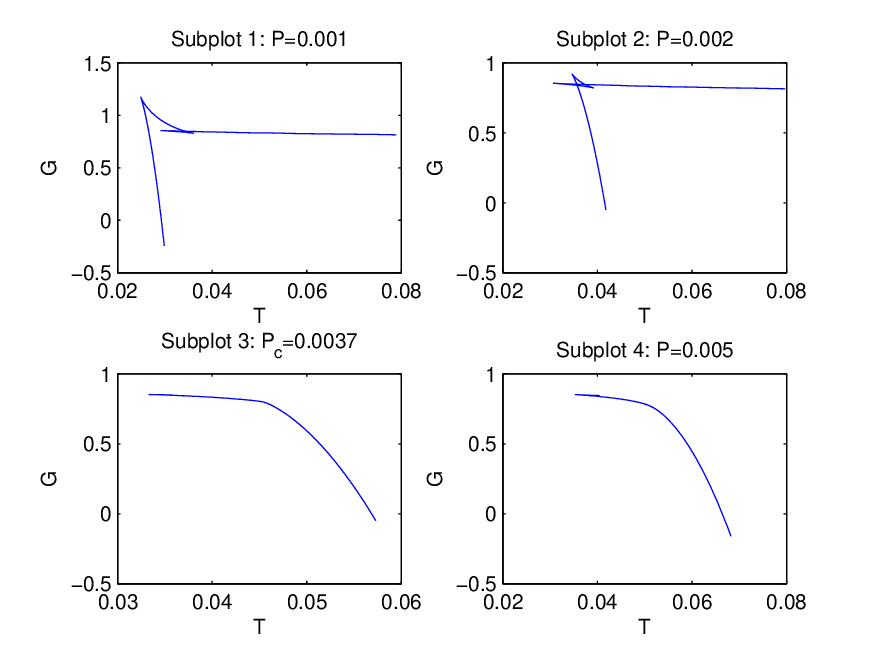}
\caption{\label{fig.5} The plots of the Gibbs free energy $G$ vs. $T$ at $q=1$, $\beta=0.5$. According to subplots 1 and 2 we have the 'swallowtail' plots with first-order phase transitions. Subplot 3 shows the second-order phase transition with $P=P_c\approx 0.0037$. Subplot 4 shows the case $P>P_c$ with non-critical behavior of the Gibbs free energy.}
\end{figure}
Subplots 1 and 2 at $P<P_c$ show first-order phase transitions (between small and large black holes for $T < T_c$) similar to liquid-gas transitions with the 'swallowtail' behaviour. In accordance with subplot 3 the second order phase transition for $P=P_c$ takes place. Subplot 4 corresponds to the case $P>P_c$, where there are not phase transitions.

The entropy $S$ vs temperature $T$ at $q=\beta=1$ is given in Fig. 6. Figure 6 (subplots 1 and 2) show that entropy is ambiguous function of the temperature and, therefore, first-order phase transitions take place.  According to  subplot 3 the second-order phase transition occurs. The critical point separates low and high entropy states. In accordance with subplot 4 there are not phase transitions at $q=\beta=1$, $P=0.005$.
\begin{figure}[h]
\includegraphics {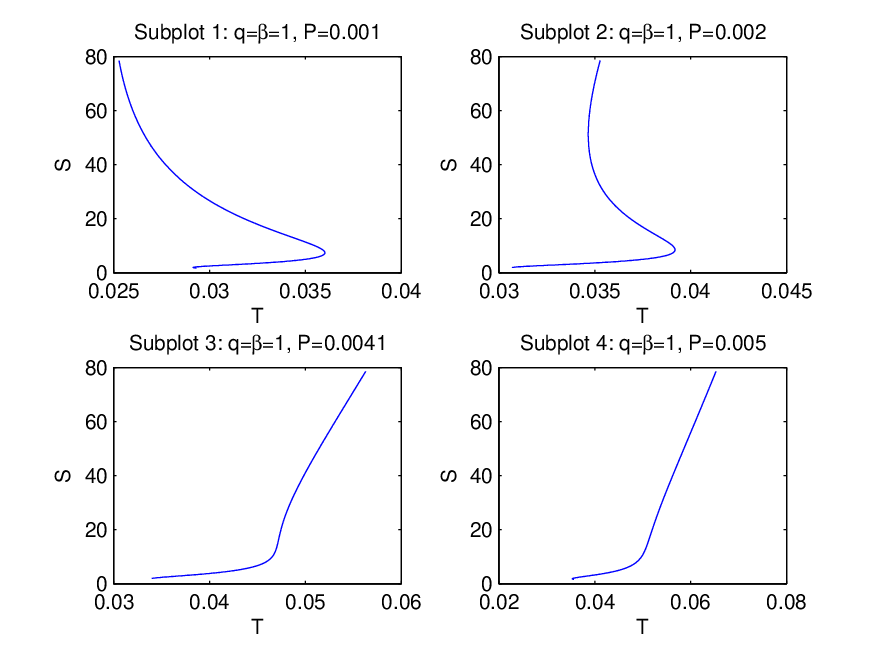}
\caption{\label{fig.6} The plots of entropy $S$ vs. temperature $T$ at $q=1,~\beta=0.5$. According to subplots 1 and 2 (in some range of $T$) entropy is ambiguous function of the temperature and first-order phase transitions occur. In accordance with subplot 3 the second-order phase transition takes place.}
\end{figure}

Let us study local stability of black holes by considering the heat capacity which is given by
\begin{equation}
C_q=T\left(\frac{\partial S}{\partial T}\right)_q=\frac{T\partial S/\partial r_+}{\partial T/\partial r_+}=\frac{2\pi r_+ T}{G_N\partial T/\partial r_+}.
\label{34}
\end{equation}
Equation (34) shows that when the Hawking temperature has an extremum the heat capacity diverges and the black hole phase transition occurs. The plot of the Hawking temperature is given in Fig. 7 for parameters $\beta=0.1,~0.3,~1$ ($l=q=1$).
\begin{figure}[h]
\includegraphics {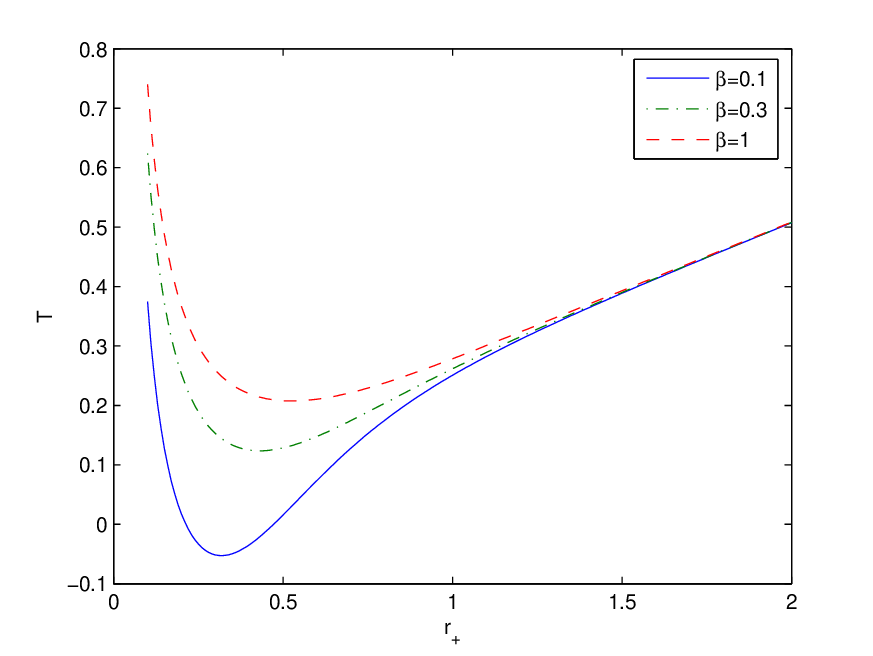}
\caption{\label{fig.7} The plots of the Hawking temperature $T$ versus horizon radius $r_+$ at $l=q=1,~\beta=0.1,~0.3,~1$. Figure shows that the Hawking temperature has a minimum.}
\end{figure}
In accordance with Fig. 7, the Hawking temperature possesses a minimum and the heat capacity diverges. Figure 7 shows that there is a region where the Hawking temperature is negative and, therefore, in this interval of event horizon radiuses, black holes do not exist. For the case $l=q=1,~\beta=0.1$, equation $T=0$ has two real roots $r_1\approx 0.213$ and $r_2\approx 0.472$.
The plot of the heat capacity (34) at $q=l=1,~ \beta=0.1$ ($G_N=1$) is depicted in Fig. 8. In accordance with Fig. 8 the heat capacity has a singularity in the point where the Hawking temperature has a minimum. The heat capacity diverges ($\partial T/\partial r_+$=0) at $r_3\approx 0.318$.
\begin{figure}[h]
\includegraphics {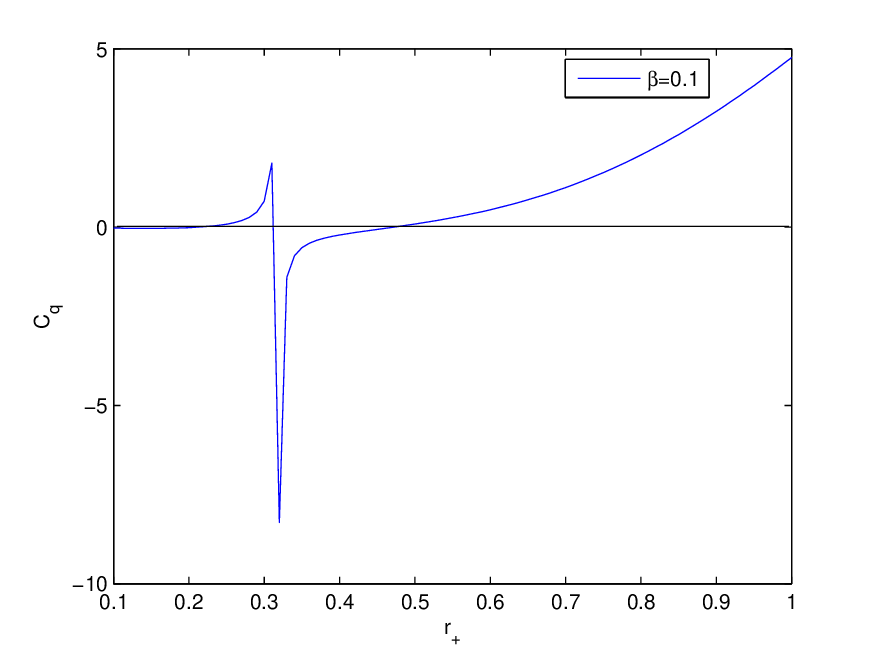}
\caption{\label{fig.8} The plot of the heat capacity $C_q$ versus horizon radius $r_+$ at $l=q=1$, $\beta=0.1$. According to the figure, the heat capacity has a singularity where the Hawking temperature possesses a minimum.}
\end{figure}
One can see from Eq. (34) that when the Hawking temperature is zero the heat capacity vanishes.
The black hole undergoes the phase transition from small black hole to large black hole in the points where the heat capacity possesses a singularity. In the region where the heat capacity is positive the black hole is stable, otherwise the black hole is unstable. At $r_2>r_+>r_1$ the Hawking temperature is negative but at $r_+>r_2$ the Hawking temperature and the heat capacity are positive and the black hole is stable.

\subsection{Reentrant phase transitions}

The reentrant phase transition firstly was observed in a nicotine/water mixture \cite{Hudson}. It was discussed also in higher dimensions \cite{Altamirano} and in spinning Kerr-AdS black holes \cite{Altamirano1}. The phenomenon of reentrant phase transition is described in multi-component fluids \cite{Narayanan}. The reentrant phase transition (zeroth-order phase transition) takes place when a system
possesses a transition from one phase to another phase and then goes back to the first phase. In this process one thermodynamic variable is changed but others remain constant. As the pressure increases from $0.001$ to $0.002$ in Fig. 8 (from panel 1 to panel 2), there will
be a large-small-large reentrant phase transition. In our model there is the global minimum of the Gibbs free energy having a jump depicted in Fig. 9 (for an example) for $P=0.002$.
\begin{figure}[h]
\includegraphics {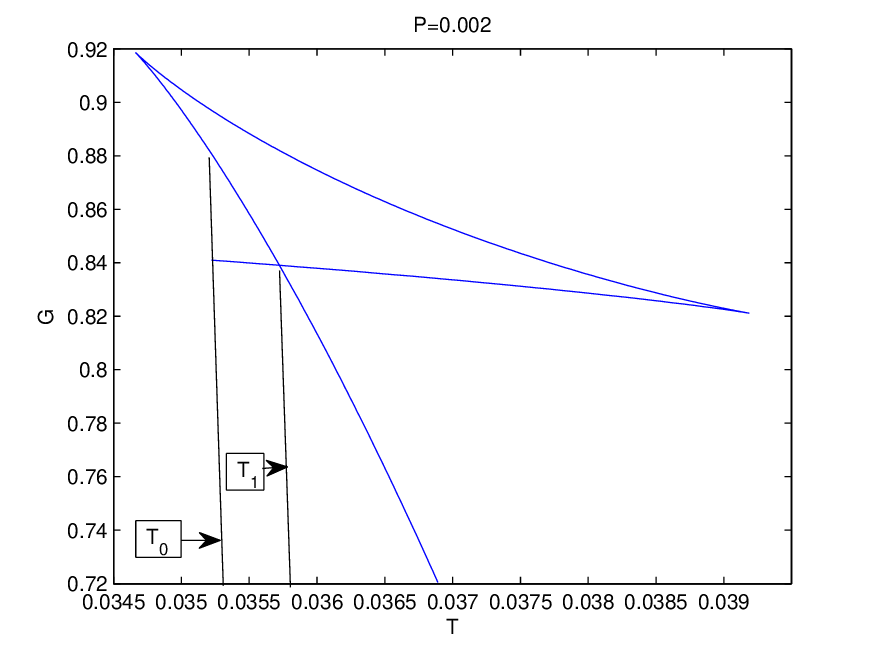}
\caption{\label{fig.9} Reentrant phase transition. There is a finite jump in the Gibbs free energy showing the zeroth-order phase transition.}
\end{figure}
When $T$ decreases, the black hole follows the lower vertical curve until $T = T_1$. Then it coincides with the upper horizontal line corresponding to small stable black holes and undergoes a first-order large-small black
hole phase transition. As $T$  decreases up to $T = T_0$, the Gibbs free energy $G$ possesses a discontinuity at its global minimum. When $T$ continues to decrease, the system goes to the  stable black holes. Thus,  the zeroth-order phase transition occurs  between small and large black holes.

\section{Joule--Thomson expansion}

During the Joule--Thomson isenthalpic expansion the enthalpy (the mass $M$) is constant. The cooling-heating phases are described by the Joule--Thomson coefficient
\begin{equation}
\mu_J=\left(\frac{\partial T}{\partial P}\right)_M=\frac{1}{C_P}\left[ T\left(\frac{\partial V}{\partial T}\right)_P-V\right]=\frac{(\partial T/\partial r_+)_M}{(\partial P/\partial r_+)_M}.
\label{345}
\end{equation}
Equation (35) shows that the Joule--Thomson coefficient is the slope of the $P-T$ function.
At the inversion temperature $T_i$ the sign of $\mu_J$ is changed, and $T_i$ can be found by equation $\mu_J(T_i)=0$. In the cooling phase ($\mu_J>0$) initial temperature is higher than inversion temperature $T_i$ and the final temperature decreases. If the initial temperature is lower than $T_i$ then the final temperature increases in accordance with the heating phase ($\mu_J<0$). Making use of Eq. (35) and  taking into account equation $\mu_J(T_i)=0$, we obtain
\begin{equation}
T_i=V\left(\frac{\partial T}{\partial V}\right)_P=\frac{r_+}{3}\left(\frac{\partial T}{\partial r_+}\right)_P.
\label{36}
\end{equation}
The inversion temperature separates cooling and heating processes. The inversion temperature
line goes through $P-T$ diagrams maxima \cite{Yaraie,Rizwan}. Equation (24) may be represented as equation of state
\begin{equation}
T=\frac{1}{4\pi r_+}+2P r_+-\frac{q^2}{4\pi\left(r_+^3+(\beta q^2)^{3/4}\right)}.
\label{37}
\end{equation}
At $\beta=0$ Eq. (37) is converted into equation of state of Maxwell-AdS black holes. From Eq. (17) and using equation $P=3/(8\pi l^2)$ one obtains
\begin{equation}
P=\frac{3}{4\pi r_+^3}\left[M(r_+)-\frac{r_+}{2}-\frac{\pi q^{3/2}}{4\sqrt{3}\beta^{1/4}}
+\frac{q^{3/2}g(r_+)}{12\beta^{1/4}}\right].
\label{38}
\end{equation}
We depicted the $P-T$ isenthalpic diagrams in Fig. 9 by taking into account Eqs. (37) and (38). Figure 10 shows that the inversion $P_i-T_i$ diagram crosses maxima of isenthalpic curves.
 \begin{figure}[h]
\includegraphics {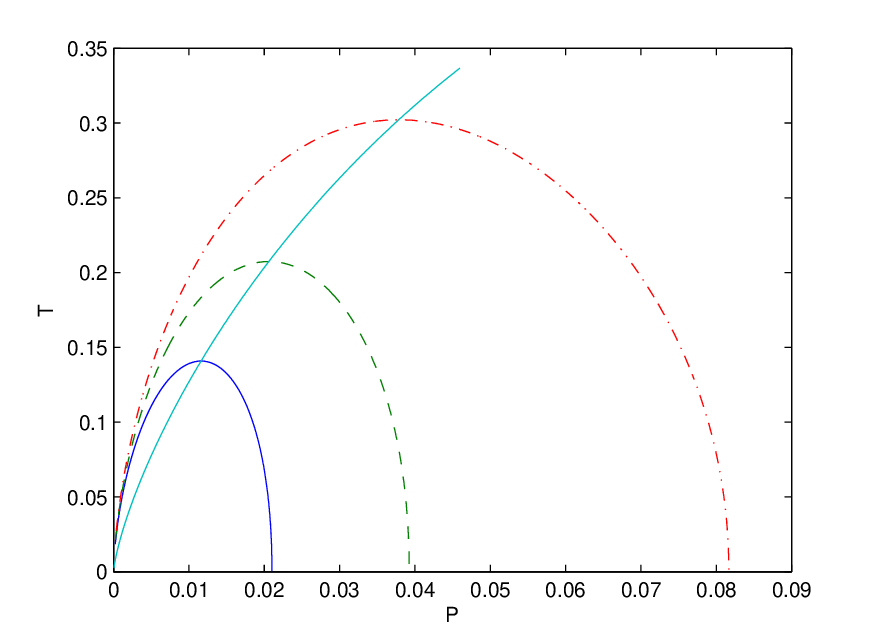}
\caption{\label{fig.10} The plots of the temperature $T$ vs. pressure $P$ for $q=30$, $\beta=0.5$.
The $P_i-T_i$ diagram goes via maxima of isenthalpic curves. The solid curve is for mass $M=90$, the dashed curve
corresponds to $M=100$, and the dashed-doted curve is for $M=110$. The inversion temperature $T_i$ vs. pressure $P_i$ ($q=30$, $\beta=0.5$) is depicted by solid line. If black hole masses are increasing the inversion temperature $T_i$ increases.}
\end{figure}
Making use of Eqs. (24), (36) and (37) we find the inversion pressure $P_i$
\begin{equation}
P_i=\frac{3q^2(2r_+^3+(\beta q^2)^{3/4})}{16\pi r_+\left(r_+^3+(\beta q^2)^{3/4}\right)^2}-\frac{1}{4\pi r_+^2}.
\label{39}
\end{equation}
By virtue of Eqs. (37) and (39) one obtains the inversion temperature
\begin{equation}
T_i=\frac{q^2(4r_+^3+(q^2\beta)^{3/4})}{8\pi (r_+^3+(q^2\beta)^{3/4})^2}-\frac{1}{4\pi r_+}.
\label{40}
\end{equation}
Substituting $P_i=0$ in Eq. (39), we find the equation for the minimum of the event horizon radius $r_{min}$
\begin{equation}
3q^2r_{min}(2r_{min}^3+(\beta q^2)^{3/4})-4\left(r_{min}^3+(\beta q^2)^{3/4}\right)^2=0.
\label{41}
\end{equation}
From Eqs. (40) and (41) at $\beta=0$, one obtains minimum of the inversion temperature corresponding to Maxwell-AdS magnetic black holes
\begin{equation}
T_i^{min}=\frac{1}{6\sqrt{6}\pi q}, ~~~~r_h^{min}=\frac{\sqrt{6}q}{2}.
\label{42}
\end{equation}
Making use of Eqs. (30) and (42) at $\beta=0$ we find the relation $T_i^{min}=T_c/2$ which corresponds to electrically charged AdS black holes \cite{Aydiner}.
With the help of Eqs. (39) and (40) we plotted $P_i-T_i$ diagrams in Fig. 6. According to Fig. 6. the inversion point increases when the black hole mass increases. The inversion diagrams $P_i-T_i$ are depicted in Figs. 10 and 11.
\begin{figure}[h]
\includegraphics {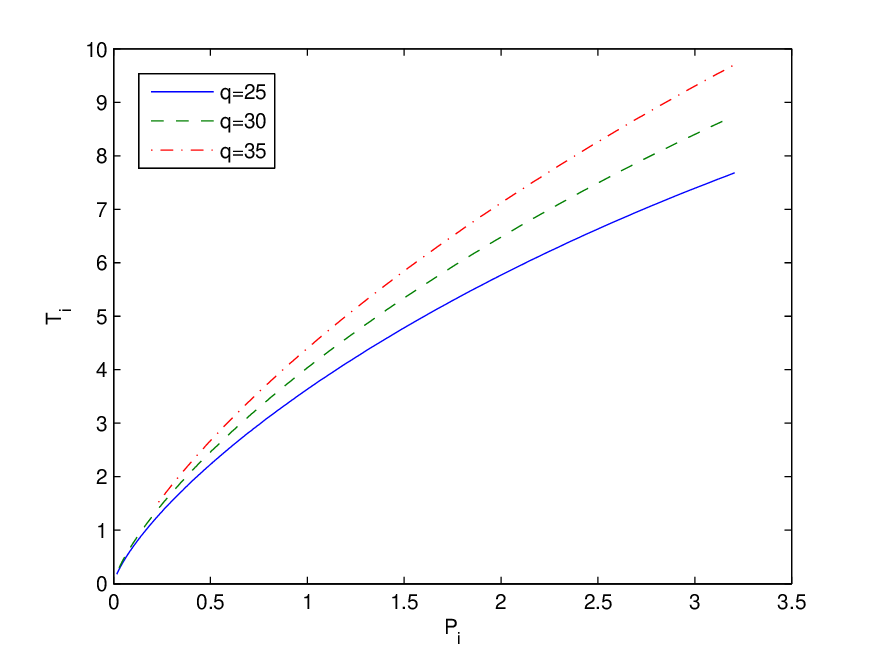}
\caption{\label{fig.11} The inversion temperature $T_i$ vs. pressure $P_i$ at $q=25$, $30$ and $35$, $\beta=0.1$. When magnetic charge $q$ increases the inversion temperature is increasing.}
\end{figure}
\begin{figure}[h]
\includegraphics {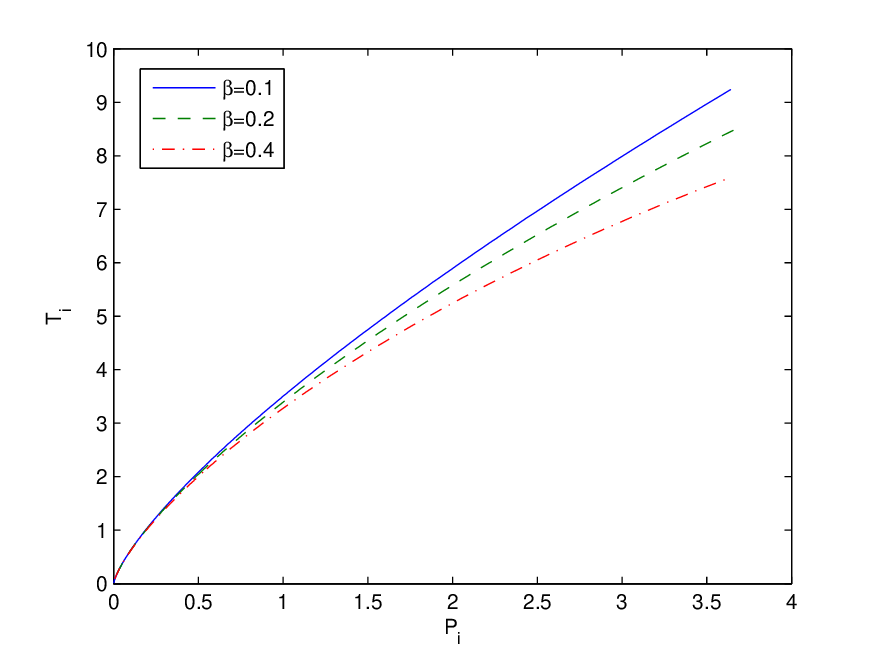}
\caption{\label{fig.12} The inversion temperature $T_i$ vs. pressure $P_i$ at $\beta=0.1$, $0.2$ and $0.4$, $q=20$. Figure shows that if the coupling $\beta$ increases the inversion temperature decreases.}
\end{figure}
According to Fig. 11 when magnetic charge $q$ increases then  the inversion temperature increases.
Figure 12 shows that when the coupling $\beta$ increases the inversion temperature decreases.
From Eqs. (35), (37) and (38) we find
\[
\left(\frac{\partial T}{\partial r_+}\right)_M=-\frac{1}{4\pi r_+^2}+2P|_M+2r_+\left(\frac{\partial P}{\partial r_+}\right)_M
+\frac{3q^2r_+^2}{4\pi[r_+^3+(q^2\beta)^{3/4}]^2},
\]
\begin{equation}
\left(\frac{\partial P}{\partial r_+}\right)_M=\frac{3}{4\pi r_+^4}\biggl[\frac{\sqrt{3}q^{3/2}\pi}{4\beta^{1/4}}-3M+r_+
-\frac{q^{3/2}g(r_+)}{4\beta^{1/4}}+\frac{q^2r_+^2}{2[r_+^3+(\beta q^2)^{3/4}]}\biggr],
\label{43}
\end{equation}
where $P|_M$ is given in Eq. (38).
Equations (35) and (43) define the Joule--Thomson coefficient as the function of the magnetic charge $q$, coupling $\beta$, black hole mass $M$ and event horizon radius $r_+$. When the Joule--Thomson coefficient is positive ($\mu_J>0$) a cooling process occurs. If $\mu_J<0$ a heating process takes place.


\section{Summary}

We obtained new magnetic black hole solution in Einstein-AdS gravity coupled to NED.
It is shown that the principles of causality and unitarity occur for any magnetic induction fields. The metric and mass functions and corrections to the Reissner--Nordstr\"{o}m solution were found. When coupling $\beta$ is increasing (at constant magnetic charge) the event horizon radius decreases. If magnetic charge increases (at constant coupling $\beta$) the event horizon radius is increasing. It was demonstrated that a spacetime singularity is present because the Kretschmann scalar is infinite at $r=0$. The black holes thermodynamics in an extended phase space with negative cosmological constant (which is a thermodynamic pressure) was studied. In this approach the mass of the black hole is the chemical enthalpy. The vacuum polarization, which is a thermodynamic quantity conjugated to coupling $\beta$, and thermodynamic potential, conjugated to magnetic charge, were obtained. We showed that the first law of black hole thermodynamics and the generalized Smarr formula take place. It was demonstrated that black hole thermodynamics is similar to the Van der Waals liquid–gas thermodynamics. We analysed the Gibbs free energy and heat capacity showing phase transitions. We have analysed zero-order, first-order, and second-order phase transitions.
The critical ratio $\rho_c$ obtained is different from the Van der Waals value $3/8$. We studied the black hole Joule--Thomson isenthalpic expansion and cooling and heating phase transitions. We found the inversion temperature which separates cooling and heating processes of black holes. There are similarities and differences in this and the past related papers. Expressions for the magnetic energy density, the mass and metric functions, the Hawking temperature, the magnetic potential and the vacuum polarization as well as the critical temperature and pressure are different for models.
It should be noted that a weak-gravity regime is released when $r$ goes to infinity. It follows from analytical expressions that as $r\rightarrow\infty$ a nonlinearity of NED disappears asymptotically.


\section{Appendix A}

The Kretschmann scalar is defined as
\begin{equation}
K(r)\equiv R_{\mu\nu\alpha\beta} R^{\mu\nu\alpha\beta}=
(f''(r))^2+\left(\frac{2f'(r)}{r}\right)^2+\left(\frac{2(f(r)-1)}{r^2}\right)^2.
\label{44}
\end{equation}
From Eq. (12) (at $G_N=1$) one finds
\[
f'(r)=\frac{2m_0}{r^2}+\frac{q^{3/2}}{6r^2\sqrt[4]{\beta q^2}}\left(g(r)+\frac{\pi}{\sqrt{3}}\right)-\frac{q^{3/2}}{r^3+\beta^{3/4}q^{3/2}}+\frac{2r}{l^2},
\]
\begin{equation}
f''(r)=-\frac{4m_0}{r^3}-\frac{q^{3/2}}{3r^3\sqrt[4]{\beta q^2}}\left(g(r)+\frac{\pi}{\sqrt{3}}\right)+\frac{4r^3+\beta^{3/4}q^{3/2}}{r(r^3+\beta^{3/4}q^{3/2})^2)}+\frac{2}{l^2},
\label{45}
\end{equation}
where
\begin{equation}
g(r)=\ln\frac{r^2-\sqrt[4]{\beta q^2}r+\sqrt{\beta} q}{(r+\sqrt[4]{\beta q^2})^2}
-2\sqrt{3}\arctan\left(\frac{1-2r/\sqrt[4]{\beta q^2}}{\sqrt{3}}\right).
\label{45}
\end{equation}
Making use of Eqs. (12) and (45)  the Kretschmann scalar versus $r$ is plotted in Fig. 12.
As $r\rightarrow 0$ the Kretschmann scalar approaches to infinity showing a spacetime singularity at $r=0$. But at small radiuses, close to the Planck length $l_P=\sqrt{G_N\hbar/c^3}$, one needs to take into account quantum effects \cite{Kruglov6}.
The Kretschmann scalar becomes constant as $r\rightarrow \infty$.
\begin{figure}[h]
\includegraphics{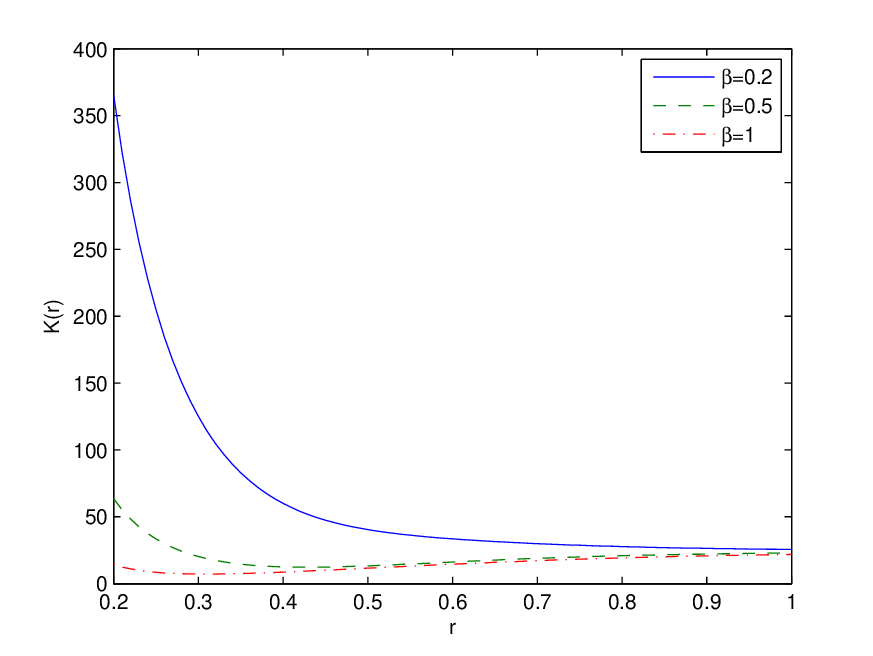}
\caption{\label{fig.12}The plot of the function $K(r)$ vs. $r$ for $q=l=1$, $m_0=0$. The solid line corresponds to $\beta=0.2$, the dashed line corresponds to $\beta=0.5$ and the dashed-dotted line corresponds to $\beta=1$. The Kretschmann scalar approaches to infinity as $r\rightarrow 0$ showing a space-time singularity at $r=0$. As $r\rightarrow \infty$ the Kretschmann scalar becomes constant.}
\end{figure}
According to Fig. 12 the curvature invariant $K(r)$ is not bounded. Figure 12 shows that as the coupling $\beta$ increases (at fixed $q$ and $l$) the Kretschmann scalar decreases.

\section{Appendix B}

The NED models are viable if causality and unitarity principles take place. The causality principle requires that a group velocity of elementary excitations over a background field does not exceed the light speed in vacuum. The unitarity principle requires that the propagator residue has to be positive. These principles lead to requirements (in our notations) \cite{Shabad}
\begin{equation}
 {\cal L}_{\cal F}\leq 0,~~~~{\cal L}_{{\cal F}{\cal F}}\geq 0,~~~~{\cal L}_{\cal F}+2{\cal F} {\cal L}_{{\cal F}{\cal F}}\leq 0,
\label{47}
\end{equation}
were ${\cal L}_{\cal F}\equiv\partial{\cal L}/\partial{\cal F}$.
Making use of of Eq. (2) we obtain
\[
{\cal L}_{\cal F}=-\frac{{\beta{\cal F}+2\sqrt[4]{2\beta {\cal F}}}}{8\pi\sqrt[4]{2\beta {\cal F}}\left(1+(2\beta{\cal F})^{3/4}\right)^2},~~~~
 {\cal L}_{{\cal F}{\cal F}}=\frac{9\beta}{32\pi\sqrt[4]{2\beta {\cal F}}\left(1+(2\beta{\cal F})^{3/4}\right)^3},
\]
\begin{equation}
{\cal L}_{\cal F}+2{\cal F} {\cal L}_{{\cal F}{\cal F}}=-\frac{2(2\beta {\cal F})^{7/4}+38\beta {\cal F}+8\sqrt[4]{2\beta {\cal F}}}{32\pi\sqrt[4]{2\beta {\cal F}}\left(1+(2\beta{\cal F})^{3/4}\right)^3}.
\label{48}
\end{equation}
Equation (47) is satisfied for $\beta> 0$ and ${\cal F}=B^2/2> 0$ i.e. for pure magnetic field.
Thus, the principles of causality and unitarity occur for any magnetic induction fields.\\
\vspace{3mm}

\textbf{Acknowledgments}
\vspace{5mm}

I wish to thank Prof. R. Mann for useful communications.

\end{document}